\documentclass[twocolumn,showpacs,preprintnumbers,amsmath,amssymb]{revtex4}
\usepackage{bm}
\usepackage{mathrsfs}
\usepackage{subfigure}
\usepackage{graphicx}

\newcommand{\tr}[1]{%
\leavevmode
\setbox0=\hbox{tr}%
\setbox1=\hbox{\scriptsize #1}%
\ifdim\wd0>\wd1 \dimen0=\wd0 \else \dimen0=\wd1 \fi
\hbox{%
  \vtop{%
    \hbox to \dimen0{\mathstrut \hfil {\rm tr} \hfil}%
      \nointerlineskip
      \vspace{-1pt}
    \hbox to \dimen0{\scriptsize \hfil#1\hfil}
}}}

\newcommand{\trs}{%
\leavevmode
\setbox0=\hbox{tr}%
\setbox1=\hbox{\scriptsize s}%
\ifdim\wd0>\wd1 \dimen0=\wd0 \else \dimen0=\wd1 \fi
\hbox{%
  \vtop{%
    \hbox to \dimen0{\mathstrut \hfil {\rm tr} \hfil}%
      \nointerlineskip
      \vspace{-1pt}
    \hbox to \dimen0{\scriptsize \hfil s\hfil}}}}
\newcommand{\trsc}{%
\leavevmode
\setbox0=\hbox{tr}%
\setbox1=\hbox{\scriptsize s,c}%
\ifdim\wd0>\wd1 \dimen0=\wd0 \else \dimen0=\wd1 \fi
\hbox{%
  \vtop{%
    \hbox to \dimen0{\mathstrut \hfil {\rm tr} \hfil}%
      \nointerlineskip
      \vspace{-1pt}
    \hbox to \dimen0{\scriptsize \hfil s,c\hfil}}}}

\begin{document}
\preprint{CYCU-HEP-10-12}
\title{
Nonet meson properties in Nambu--Jona-Lasinio model with dimensional 
versus cutoff regularization
}

\author{
T. Inagaki
}
\affiliation{
Information Media Center, Hiroshima University,
Higashi-Hiroshima, Hiroshima
739-8521, Japan}
\author{
D. Kimura 
}
\affiliation{
Learning Support Center, Hiroshima Shudo University,
  Hiroshima, 731-3195, Japan
}
\author{
H. Kohyama
}
\affiliation{
Institute of Physics, Academia Sinica,
 Taipei 115, Taiwan and\\
 Physics Division, National Center for Theoretical Sciences,
 Hsinchu 300, Taiwan and\\
 Department of Physics, Chung-Yuan Christian University, Chung-Li 32023, Taiwan
 }

\author{
A. Kvinikhidze
}
\affiliation{
A. Razmadze Mathematical Institute of Georgian Academy of Sciences,\\
  M. Alexidze Str. 1, 380093 Tbilisi, Georgia}

\date{\today}

\begin{abstract}
Nambu--Jona-Lasinio (NJL) model with Kobayashi-Maskawa-'t Hooft (KMT) 
term is one of low energy effective theory of QCD which includes the $U_A(1)$
anomaly. We investigate nonet meson properties in this model with three
flavors of quarks. 
We employ two type of regularizations the dimensional and 
sharp cutoff ones. 
The model parameters are fixed phenomenologically for 
each regularization. Evaluating the kaon decay constant, the $\eta$ meson 
mass and the topological susceptibility, we show 
the regularization dependence of the results and discuss the applicability of 
the NJL model.
\end{abstract}


\maketitle

\section{INTRODUCTION}
The unique QCD Lagrangian is found by imposing the local $SU_c(3)$ gauge symmetry, 
Lorentz invariance, locality and renormalizability in four space-time dimensions. 
It is believed to be a fundamental theory of strong interaction between quarks and gluons. 
One of the features of QCD dynamics is ``asymptotic freedom'' which
legitimates perturbation theory at short distances. 
Unfortunately non-perturbative effects cannot be avoided in the
confinement phase which 
takes place at low energy scale, i.e. where the QCD coupling is not small.

Nambu--Jona-Lasinio (NJL) model \cite{NJL}
is a well-known and often used low energy effective theory of QCD
\cite{Vogl:1991qt, Klevansky:1992qe, Hatsuda:1994pi}. 
Nambu and Jona-Lasinio have introduced a four-fermion interaction to
describe the attractive force 
between fermions. In this model,  chiral symmetry is spontaneously broken by non-vanishing 
expectation value for a composite operator constructed by the fermion and anti-fermion fields 
and the fermion mass is dynamically generated. The NJL model and its generalizations 
are extremely useful in the study of the light meson properties at low energy.

The four-fermion interaction is a dimension six operator in four space-time dimensions
therefore the model is non-renormalizable and depends on the
regularization procedure.
In order to regularize fermion loop integrals one usually
introduces a momentum scale $\Lambda$ to cutoff integration momenta higher than $\Lambda$. 
Another regularization, the dimensional one, is an analytic
regularization; one calculates 
loop integrals as analytic functions of the space-time dimensions and uses them for the value
of  the space-time dimensions less than four \cite{Krewald:1991tz,
Inagaki:1994ec, Jafarov:2004jw, Inagaki:2007dq, Fujihara:2008ae}.

In the present paper we study NJL model in the dimensional regularization and compare
the results with ones obtained in the sharp cutoff regularization. 
NJL model in both regularizations describes well the dynamical 
breaking of chiral symmetry and the $\pi$ and $\sigma$ meson properties 
in vacuum \cite{Inagaki:2007dq}. Yet at finite density (especially when
the cutoff scale 
is close to the Fermi momentum) important contributions to the quark
loop integrals are 
dropped in the cutoff regularization scheme \cite{footnote},
while dimensional regularization leads to 
results consistent with QCD even in the region of asymptotic freedom,  
therefore we find a strong
regularization dependence in the extended NJL model with an attractive force in the color 
anti-triplet channel for a large chemical potential \cite{Fujihara:2008ae}. 
Such a high density state 
may be realized in astrophysical objects. It is expected that the model can be tested by 
observing the structure of dense stars.

Non-negligible regularization dependence is observed even when 
considering the system at $T=\mu=0$. 
The cutoff scale is usually taken to be lower than $\eta'$ meson mass ($m_{\eta'}\simeq
958$MeV) 
to reproduce light meson properties \cite{Hatsuda:1994pi, Lutz:1992dv, Rehberg:1995kh}.
Then the $m_{\eta'}$ in the low energy effective theory is not well-defined quantity.
$\Lambda$ is phenomenologically fixed and it is considered as a scale above which 
the effective model may lose its validity.
Furthermore, the cutoff regularization 
may break some symmetry of the Lagrangian. The cutoff regularization 
may cause some unexpected effects in the system where the strange quark and $U_A(1)$ anomaly 
play an important role. Therefore we launch a plan to study the three flavor system 
with $U_A(1)$ anomaly by using the dimensional regularization.

In the present paper the extended NJL model including the
Kobayashi-Maskawa-'t Hooft (KMT) term \cite{Kobayashi:1970ji, 'tHooft:1976fv} 
is regarded as a low energy effective theory of QCD with $U_A(1)$ anomaly.
We consider three-flavor light quarks, $u,\ d$ and $s$, and investigate the nonet 
meson properties. In Sec.~II. we introduce the model Lagrangian and briefly 
review regularization schemes. In Sec.~III we calculate the meson masses, 
decay constants and topological susceptibility in the leading order of 
$1/N_c$ expansion. The results depend on the regularization parameters. 
In Sec.~IV we evaluate the kaon decay constant, the $\eta$ meson mass, 
$m_\eta$, and the topological susceptibility $\chi$ as a function of the
space-time dimension $D$ associated with loop integrals. 
In Sec.~V the dependence of physical quantities
on the current up-quark mass, $m_u$, is studied.
After phenomenologically fixing the model parameters, 
we discuss the two regularizations
dependence of the results. 
This paper shows that the model Lagrangian is not
satisfactory in either of regularizations.
Some concluding remarks are given in Sec.~VI.

\section{NJL MODEL WITH $U_A(1)$ ANOMALY}
\label{njl_model}

The NJL model is one of the simplest models to describe the dynamical
symmetry breaking. It is often used to study the light meson properties
at low energy scale. Since the $U_A(1)$ anomaly induces the mass
difference between $\eta$ and $\eta'$ mesons, the NJL model should
be extended to include the contribution from the $U_A(1)$ anomaly
in evaluating the nonet meson system constructed 
out of the three flavor light 
quarks.

\subsection{Model set up}
Kobayashi and Maskawa have introduced an interaction written
in the determinant of the composite operator constructed by 
quark and anti-quark fields in the nonet meson system
\cite{Kobayashi:1970ji}.
In QCD $\theta$-term can be transformed to the determinant term 
by $U_A(1)$ transformation. Thus we can include the contribution 
of the $U_A(1)$ anomaly through the determinant term in the NJL 
model. In the present paper we consider the four- and six-fermion
interaction invariant under $SU_L(3)\otimes SU_R(3)$ global
flavor symmetry and start with the Lagrangian,
\begin{equation}
 \mathcal{L}_{\mathrm{NJL}} = \sum_{i,j=1}^3
 \bar{q}_i\left( i \partial\!\!\!/
  - \hat{m}\right)_{ij}q_j + \mathcal{L}_4 + \mathcal{L}_6 ,
\label{LNJL}
\end{equation}
where
\begin{align}
 \mathcal{L}_4 &= G \sum_{a=0}^8 \left[
  \left( \sum_{i,j=1}^3 \bar{q}_i\lambda_a q_j\right)^2
  + \left( \sum_{i,j=1}^3\bar{q}_i\,i \gamma_5 \lambda_a q_j \right)^2
  \right] ,
\label{L_4} \\
 \mathcal{L}_6 &= -K \left[ \det\bar{q}_i (1-\gamma_5) q_j 
 +\text{h.c.\ } \right] ,
\label{L_6}
\end{align}
the subscripts ($i$,$j$) 
are the flavor indices,
$\hat{m}$ expresses the current quark mass matrix,
$\lambda_a$ are the Gell-Mann matrices in the flavor space,
$G$ and $K$ represent the effective coupling constants
for four- and six-fermion interaction, respectively. 
The determinant in $\mathcal{L}_6$ 
concerns the matrix elements labeled by the flavor indices, $ij$. 
The order of the coupling constants are supposed 
to be $GN_c \simeq O(1)$, $KN_c^2 \simeq O(1)$ therefore we work 
in the framework of the $1/N_c$ expansion.
In this paper we do not care about the flavor 
mixing and set the mass matrix to 
have a diagonal form,
$\hat{m}=\mbox{diag}(m_u,m_d,m_s)$. The current quark 
mass explicitly breaks the global $SU(3)$ flavor symmetry.
Below we consider the $SU(2)$ isospin symmetric case and
take $m_u=m_d$ for simplicity.

The chiral condensates $\langle \bar{u}u \rangle$, 
$\langle \bar{d}d \rangle$ and $\langle \bar{s}s \rangle$
generate the constituent quark masses, $m^*_u$, $m^*_d$
and $m^*_s$ inside mesons. To evaluate the constituent quark 
mass we solve the gap equations which are derived by 
differentiating the thermodynamic potential with respect
to $m^*_u$, $m^*_d$ and $m^*_s$.
The gap equation is obtained in the leading order of $1/N_c$ 
expansion
\cite{Vogl:1991qt, Klevansky:1992qe, Hatsuda:1994pi},
\begin{align}
m_u^{*} &=m_d^{*}=  
m_u + 4 G (i\, \trsc S^u) + 2 K (i\, \trsc S^d)(i\, \trsc S^s),
\label{gap_u} \\
m_s^{*} &= m_s + 4 G (i\, \trsc S^s) + 2 K (i\, \trsc S^u)(i\, \trsc S^d),
\label{gap_s}
\end{align}
where the symbol $\trsc$ stands for the trace in
$\underline{\rm s}$pinor and $\underline{\rm c}$olor indices.
$\trsc S^i$ represent the chiral condensates,
$\displaystyle -i \trsc S^u \equiv \langle \bar{u}u \rangle$ and
$\displaystyle -i \trsc S^s \equiv \langle \bar{s}s \rangle$,
which are given by the trace of the quark propagator inside 
mesons,
\begin{align}
 -i\, \trsc S^i &= N_c \trs \int \!\! \frac{d^Dp}{i(2\pi)^D}\,S^i(p), 
\label{trace} \\
 S^i(p) &\equiv \frac{1}{p\!\!\!/-m_i^*+i\epsilon} .
\nonumber
\end{align}
Here we indicate the space-time dimension for internal quark 
fields by $D$.
Below we omit the subscripts in $\trsc$ for notational simplicity.

\subsection{Regularization schemes}
The fermion loop integral in  Eq.(\ref{trace}) is divergent in four
space-time dimensions to obtain a finite result we have to regularize it. 
The four- and six-fermion interaction are written 
in terms of the dimension six and nine 
operators, respectively. Thus the operators in $\mathcal{L}_4$ and 
$\mathcal{L}_6$ are irrelevant in four dimensions. It means that the
results depend on regularization procedures. Here we use 
two different procedures;  
one is the three-momentum sharp cutoff 
regularization and the other is the dimensional regularization.

In the three-momentum sharp cutoff method, we cut off the space 
momentum component integral 
above the scale, $\Lambda$,
\begin{align}
 \int \!\frac{d^Dp}{(2\pi)^D} 
  \rightarrow
 \int \frac{dp_0}{2\pi}  \int^{\Lambda} \!\! \frac{d^3p}{(2\pi)^3}.
\end{align}
In the dimensional regularization scheme, we regularize the divergent
integral with the help of analytic continuation of the integral as a 
function of the space-time dimension $D$ to a
non-integer value less than four,
\begin{align}
 \int \!\frac{d^Dp}{(2\pi)^D} 
  \rightarrow
 \frac{2\,(4\pi)^{-D/2}}{\Gamma(D/2)} \int_0^{\infty} \!\!dp \,p^{D-1}.
\end{align}
This is a kind of an analytic regularization. We regard the space-time 
dimensions $D$ in the fermion loop integral as one of the parameters 
of the effective model of QCD. The dimensional regularization is 
applied to momentum integrals only for internal fermion lines.

The common parameters of the models considered here are the coupling constants
$G$ and $K$, the current quark masses $m_u(=m_d)$, $m_s$. In the cutoff scheme,
the cutoff scale $\Lambda$ is one more parameter.
On the other hand in the dimensional regularization, we consider the space-time
dimension $D$ as one of the model parameters. In this case 
we have to introduce one more parameter, the renormalization scale $M_0$, 
to obtain results with the correct mass dimension. Thus the parameters
in these two regularization methods are aligned as follows,
\begin{quote}
Cutoff:\hspace{0.2cm} $G$, $K$, $m_u(=m_d)$, $m_s$, $\Lambda$,\\
Dimensional:\hspace{0.2cm} $G$, $K$, $m_u(=m_d)$, $m_s$, $D$, $M_0$.\\
\end{quote}
All the parameters should be fixed phenomenologically.

\section{MESON MASS AND DECAY CONSTANT}
\label{meson}
In this section we shall evaluate the properties 
of the nonet meson system.
Here we calculate the meson mass, meson decay 
constant and topological susceptibility 
in the two regularizations.

\subsection{$\pi$ and $K$ masses}
First, we consider pion and kaon. The masses of these mesons are
obtained by observing the pole structure in their propagators. 
Employing the random-phase approximation (RPA) and the 
$1/N_c$ expansion, the meson propagators are given by 
\cite{Klevansky:1992qe, Hatsuda:1994pi}  
\begin{equation}
\Delta_P(k^2) = \frac{2K_\alpha}{1-2K_\alpha\Pi_{P}(k^2)} 
+\mbox{O}({N_c}^{-1}),
\label{pro_P}
\end{equation}
where the index $\alpha$ denotes the isospin channel and $P$ 
stands for the meson species, $\pi$ and $K$.
The flavor-dependent effective couplings $K_\alpha$ are defined by
\begin{align}
K_3 &\equiv G + \frac{1}{2} K i\,{\rm tr} S^s, \quad {\rm for}\,\,\,\, \pi^0,
\label{K_3}\\
K_6 &\equiv G + \frac{1}{2} K i\,{\rm tr} S^u, \quad {\rm for}\,\,\,\, K^0, \bar{K}^0.
\label{K_6}
\end{align}
In the leading order of the $1/N_c$ expansion the self-energy for 
each meson, $P$, is given by
\begin{align}
&\Pi_{P}(k^2) \delta_{\alpha\beta} \nonumber \\
&= \int \frac{d^D p}{i(2\pi)^D}
  {\rm tr}\bigl[ \gamma_5 T_\alpha  S^i(p+k/2) \gamma_5 
  T_\beta^\dagger S^j(p-k/2)\bigr],
\label{Pi_P}
\end{align}
where the trace runs over flavor, spinor and color indices.
The $SU(3)$ matrices, $T_\alpha$,
corresponding to different channels are,
$T_3=\lambda_3$ for $\pi^0$, $T_6=(\lambda_6+i\lambda_7)/\sqrt{2}$ 
for $K^0$ and $T_6^\dagger$ for $\bar{K}^0$. Thus the self-energy for $\pi^0$,
$K^0$ and $\bar{K}^0$ read
\begin{eqnarray}
  \Pi_{\pi}(k^2=m_\pi^2) &=& 2 \Pi_5^{uu}(k^2=m_\pi^2), 
\label{Pi_pi} \\
  \Pi_{K}(k^2=m_K^2) &=& 2\Pi_5^{sd}(k^2=m_K^2),
\label{Pi_K}
\end{eqnarray}
where $\Pi_5^{ij}(k^2)$ is the loop integral
\begin{eqnarray}
\Pi_5^{ij}(k^2) &=& \! \int \!\frac{d^D p}{i(2\pi)^D}
  \, {\rm tr}\bigl[ \gamma_5 S^i(p+k/2) \gamma_5 
   S^j(p-k/2)\bigr]  \nonumber \\
&=& \frac12 \left( \frac{i\,{\rm tr}S^i}{m_i^*} 
   + \frac{i\,{\rm tr}S^j}{m_j^*} \right) \nonumber \\
&& +\frac12 \bigl[ k^2 - (m_i^* - m_j^*)^2 \bigr] \,I_{ij}(k^2) ,
\label{Pi_5}
\end{eqnarray}
with
\begin{align}
I_{ij}(k^2)
= \int \! \frac{d^D p}{i(2\pi)^D} \,
         \frac{\trs1}{ ( p^2-m_i^{* \, 2} ) \bigl[ (p-k)^2-m_j^{*\,2} \bigr] }.
\label{I_ij}
\end{align}
Here we set $\trs1$ to be $4$ in the cutoff 
and $2^{D/2}$ in the dimensional
regularization schemes respectively.

Substituting the solution of the gap equations (\ref{gap_u}) 
and (\ref{gap_s}) in the self-energy expression for $\pi^0$,
$K^0$ and $\bar{K}^0$ and evaluating the pole structure of
the denominator in Eq.(\ref{pro_P}), we obtain the on-shell
conditions for pion and kaon masses,
\begin{align}
0&=\frac{m_u}{m_u^*}
 -2 K_3 k^2 I_{uu}(k^2)|_{k^2=m_\pi^2},
\label{m_pi} \\
0&=1-\frac{m_s^*-m_s}{2m_u^*}-\frac{m_u^*-m_u}{2m_s^*}
\nonumber \\
&-2G\left(\frac{i{\rm tr}S^u-i{\rm tr}S^s}{m_u^*} 
+\frac{i{\rm tr}S^s-i{\rm tr}S^u}{m_s^*} \right) 
\label{m_K} \\
&-2 K_6 \bigl[ k^2-(m_s^*-m_u^*)^2 \bigr] \, I_{us}(k^2)|_{k^2=m_K^2} .
\nonumber
\end{align}
These equations are used to determine the values of the constituent quark masses
$m_u^*$ and $m_s^*$. Because of the $SU(2)$ isospin symmetry the equations for the charged
pion and kaon cases are the same as for the neutral ones.
The QED effect is also important for the mass differences between
the neutral and  charged mesons.
It should be noted that Eq.(\ref{m_K}) becomes Eq.(\ref{m_pi})
in the limit $m_s \rightarrow m_u$ because the $SU(3)$ flavor symmetry is 
restored in this limit.

\subsection{$\pi$ and $K$ decay constants}
Next, we consider the pion and kaon decay constants. The decay 
constant $f_P$ is defined by the matrix element of the axial 
current between meson and vacuum states,
\begin{align}
&ik_\mu f_{P} \delta_{\alpha\beta} \nonumber \\
&= - M_0^{4-D} \!\int \! \frac{d^D p}{(2\pi)^D}
  {\rm tr}\left[\gamma_\mu \gamma_5 \frac{T_\alpha}{2} 
  S^i g_{Pqq}(0) \gamma_5 T_\beta^\dagger S^j \right],
\end{align}
where we introduce a renormalization scale $M_0$ to define
the decay constant with a correct mass dimension, dim$(f_P)=1$
for dimensional scheme. The trace runs over flavor, spinor and color 
indices. The meson-to-quark-quark coupling $g_{Pqq}$ is
given by
\begin{equation}
g_{Pqq}(k^2)^{-2} = \left. M_0^{4-D}\frac{\partial \Pi_P(k^2)}
 {\partial k^2} , \right . 
\label{g_Pqq}
\end{equation}
Inserting Eqs.\ (\ref{Pi_pi}) and (\ref{Pi_K})
into Eq.\ (\ref{g_Pqq}), we obtain the pion and kaon 
decay constants,
\begin{align}
f_\pi^2 M_0^{D-4} = & m_u^{*2} I_{uu}(0) ,
\label{F_pi} \\
f_K^2 M_0^{D-4}= &\frac1{J_{us}(0)} \Bigl[m_u^* I_{us}(0) 
 + \trs1 \cdot N_c (m_s^* -m_u^*) 
\nonumber \\
& \times \int_0^1 \! dx \int \! \frac{d^D p}{i(2\pi)^D}
 \frac{x}{ ( p^2 - L_{us}(0) + i\epsilon )^2} \biggr]^2 ,
\label{F_K}
\end{align}
where $J_{us}$ is defined by
\begin{align}
J_{us}(k^2) = &I_{ij}(k^2) + 2N_c (m_s^* - m_u^*)^2 \nonumber \\
 &\times \int_0^1 \! dx \int \! \frac{d^D p}{i(2\pi)^D} 
 \frac{x(1-x)}{ ( p^2 - L_{us}(k^2) + i\epsilon )^3} .
\label{J_us}
\end{align}
$L_{ij}$ is defined in the appendix (see, Eq.(\ref{L_ij})).
As confirmed in the case with $m_{\pi}$ and $m_K$, Eq.(\ref{F_K}) corresponds to
Eq.(\ref{F_pi}) in the $m_s \rightarrow m_u$ limit due to the restoration of
the $SU(3)$ flavor symmetry.

\subsection{$\eta$ and $\eta'$ mesons}
As is well known, the octet state, $\eta_8$, and the singlet 
state, $\eta_0$, are mixed in the real world. Thus the eigen-states, 
$\eta$ and $\eta'$, with diagonal mass matrix are 
described as mixed states of $\eta_8$ and $\eta_0$.
In the RPA the propagator of the $\eta -\eta'$ system
is given by \cite{Klevansky:1992qe, Hatsuda:1994pi}
\begin{equation}
\Delta^+(k^2) = 2K^+ [1-2K^+ \Pi(k^2)]^{-1} ,
\label{pro^+}
\end{equation}
where $K^+$ and $\Pi$ are the $2\times2$ matrices
\begin{eqnarray}
K^+ &=& \left(
\begin{array}{cc}
 K_{00} & K_{08} \\
 K_{80} & K_{88}
\end{array}
\right) ,
\label{K^+} \\
\Pi &=& \left(
\begin{array}{cc}
 \Pi_{00} & \Pi_{08} \\
 \Pi_{80} & \Pi_{88}
\end{array}
\right) ,
\label{Pi}
\end{eqnarray}
with
\begin{align*}
&K_{00}=G - \frac{1}{3} K(i\,{\rm tr} S^s + 2i \,{\rm tr}S^u),
\\ 
&K_{88}=G - \frac{1}{6} K(i\,{\rm tr} S^s - 4i\,{\rm tr}S^u),
\\ 
&K_{08}=K_{80}=- \frac{\sqrt{2}}{6} 
 K(i\,{\rm tr} S^s - i\,{\rm tr}S^u),
\end{align*}
and
\begin{align*}
&\Pi_{00}(k^2)=\frac{2}{3}\left[ 2\Pi_5^{uu}(k^2)+\Pi_5^{ss}(k^2) \right],
\\ 
&\Pi_{88}(k^2)=\frac{2}{3}\left[ \Pi_5^{uu}(k^2)+2\Pi_5^{ss}(k^2) \right],
\\ 
&\Pi_{08}(k^2)=\Pi_{80}(k^2)=\frac{2\sqrt{2}}{3} \left[ \Pi_5^{uu}(k^2)
                 -\Pi_5^{ss}(k^2) \right].
\end{align*}
To obtain the $\eta$ and $\eta'$ meson masses we diagonalize the
inverse propagator via an orthogonal transformation
\cite{Hatsuda:1994pi, Rehberg:1995kh}. 
Thus the $\eta$ and $\eta'$ meson masses are given by the solution
to the following equations,
\begin{align}
A(k^2)+C(k^2)-\sqrt{\{A(k^2)-C(k^2)\}^2 + 4\{B(k^2)\}^2 } &=0 
\nonumber \\ 
{\rm for}\ k^2=m_\eta^2 ,  
\label{m_eta}  \\ 
A(k^2)+C(k^2)+\sqrt{\{A(k^2)-C(k^2)\}^2 + 4\{B(k^2)\}^2 } &=0 
\nonumber \\
{\rm for}\ k^2=m_{\eta'}^2 , 
\label{m_eta'}
\end{align}
where 
\begin{eqnarray*}
A(k^2) &=& K_{88} - 2\Pi_{00}(k^2) \det K^+ , \\
B(k^2) &=& -K_{08} - 2\Pi_{08}(k^2) \det K^+ , \\
C(k^2) &=& K_{00} - 2\Pi_{88}(k^2) \det K^+ .
\end{eqnarray*}
The mixing angle $\theta_\eta$ is found to be
\begin{equation}
\tan(2\theta_\eta) = \frac{2B(k^2)}{A(k^2)-C(k^2)} .
\end{equation}

The off-diagonal matrix elements in Eq.(\ref{pro^+}) disappear and the mixing
angle $\theta_\eta$ vanishes in  the limit $m_s \rightarrow m_u$.
In this limit Eq.(\ref{m_eta}) coincides with Eqs.(\ref{m_pi}) and
 (\ref{m_K}) and the mass  spectrum for the octet mesons degenerates. 
Since the $U_A(1)$ anomaly breaks the degeneracy between the octet and 
the singlet mesons, a different on-shell condition is obtained from
Eq.(\ref{m_eta'}).

\subsection{Topological susceptibility}
To compare the QCD axial current with the NJL axial current
we introduce the topological charge density 
$Q(x)$ \cite{Hatsuda:1994pi},
\begin{align}
Q(x)\equiv\frac{g^2}{32\pi^2}F^a_{\mu\nu} \tilde{F}^{a\mu\nu}
= 2K\, {\rm Im} [\det \bar{q}(1-\gamma_5)q],
\end{align}
where $g$ is the strong coupling constant of QCD and $F^a_{\mu\nu}$ 
is the field strength for gluons. The topological susceptibility 
$\chi$ is defined by the correlation function between the topological 
charge densities at different points,
\begin{equation}
\chi =\, \int d^4x \langle 0| TQ(x)Q(0) 
 |0\rangle_{\rm connected} .
\end{equation}
It describes some global feature of QCD dynamics.
In the leading order of $1/N_c$ expansion it is given by 
\cite{Fukushima:2001prc}
\begin{align}
\chi
=\, &\frac{4K^2}{M_0^{D-4}}(i{\rm tr}S^u)^2 \left[ (i{\rm tr}S^u)(i{\rm tr}S^s) 
 \left(\frac{2i{\rm tr}S^s}{m_u^*} + 
 \frac{i{\rm tr}S^u}{m_s^*} \right) \right.
\nonumber \\
&+\left\{ \frac{1}{\sqrt{6}} (2i{\rm tr}S^s + i{\rm tr}S^u)
 \bigl( \Pi_{00}(0), \Pi_{08}(0) \bigr) \right.
\nonumber \\
&+\left. \frac{1}{\sqrt{3}} (i{\rm tr}S^s - i{\rm tr}S^u)
 \bigl( \Pi_{08}(0), \Pi_{88}(0) \bigr)
\right\} \Delta^+(0)
\nonumber \\
&\times\left\{ \frac{1}{\sqrt{6}} (2i{\rm tr}S^s + i{\rm tr}S^u)
\left(
\begin{array}{c}
 \Pi_{00}(0) \\
 \Pi_{08}(0) 
\end{array}\right) \right.
\nonumber \\
&+\left. \frac{1}{\sqrt{3}} (i{\rm tr}S^s - i{\rm tr}S^u)
\left( \left.
\begin{array}{c}
 \Pi_{08}(0) \\
 \Pi_{88}(0)
\end{array} \right) \right\} \right] .
\label{chi}
\end{align}

We compare this result with the one obtained by the lattice QCD.

\section{PARAMETER SETTING IN THE DIMENSIONAL REGULARIZATION}

In this section we discuss the physical scale and the parameter
setting for the model with the dimensional regularization.
In the study with the help of the NJL model, parameters are usually determined by 
fitting the physical quantities $\{m_{\pi},\, f_{\pi},\, m_{K},\, X\}$, 
where there are several choices for $X$. Here we fit the parameters by 
choosing $X=m_{\eta^{\prime}}$.
We reproduce these four observables  in the dimensional regularization
scheme without fixing two of model parameters, $m_u$ and $D$.
As for the up quark mass, $m_u$, we fix it by hand and test several values.
The dimension, $D$, is kept to be a free parameter.
Then we calculate some meson characteristics as functions of $D$.
Here we evaluate $D$-dependence of some observed physical quantities:
the kaon decay constant $f_K$, $\eta$ meson mass $m_{\eta}$ and
topological susceptibility $\chi$.

\subsection{Parameter setting}
\begin{figure}[!h]
  \begin{center}
    \includegraphics[height=2.0in,keepaspectratio]
    {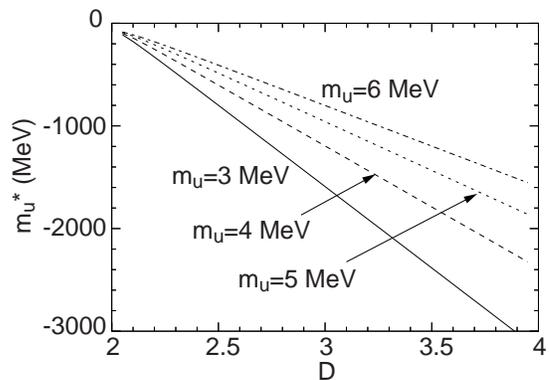} 
  \end{center}
  \vspace{-0.5cm}
  \caption{The solution of the gap equation for $m_u^*$
   as a function of $D$ for $m_u=3,4,5$ and $6$MeV.}
  \label{Mu}
\end{figure}

\begin{figure}[!h]
  \begin{center}
    \includegraphics[height=2.0in,keepaspectratio]
    {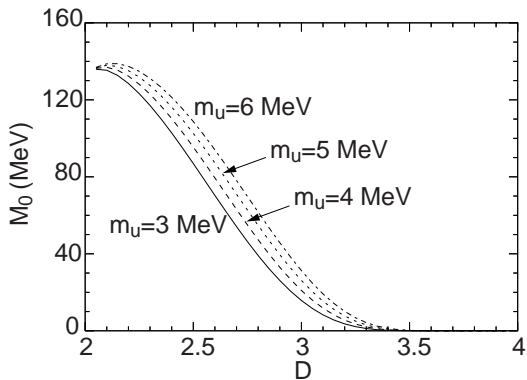} 
  \end{center}
  \vspace{-0.5cm}
  \caption{The renormalization scale $M_0$
   as a function of $D$ for $m_u=3,4,5$ and $6$MeV.}
  \label{scale}
\end{figure}

The model has $6$ parameters $m_u$, $m_s$, $G$, $K$, $M_0$ and $D$,
as is already mentioned in Sec.~\ref{njl_model}. Here we set $m_u(=m_d)$ 
at $3$, $4$, $5$ and $6$MeV, and study the region, $2<D<4$, in which
an UV stable fixed point appears for $G$.
Following the previous study \cite{Hatsuda:1994pi}, we
fit the other 4 parameters \{$M_0$, $G$, $K$, $m_s$\} by using the
measured physical observables \cite{PDG},
\begin{align}
\begin{array}{c}
m_{\pi}=138{\rm MeV},\ m_K=495 {\rm MeV}, \\ 
f_{\pi}=92 {\rm MeV},\ m_{\eta^{\prime}}=958 {\rm MeV}.
\end{array}
\label{input}
\end{align}
These quantities are calculated by analyzing the Eqs.\ (\ref{m_pi}),\
 (\ref{m_K}),\ (\ref{F_pi}) and (\ref{m_eta'}), respectively.
We solve these equations under the constraints imposed by the gap equations
for $m_u^*$ and $m_s^*$. Eliminating the term
$K i{\rm tr}S^s$ 
from Eqs.~(\ref{gap_u}) and (\ref{m_pi}), we
obtain $m_u^*$ as a function of $m_u$ and $D$,
\begin{equation}
0=\frac{m_u}{m_u^*} - \frac{1}{2} \frac{m_u^*-m_u}{i{\rm tr}S^u}
k^2 I_{uu}(k^2)|_{k^2=m_\pi^2} .
\label{sol_gap_u}
\end{equation}
We numerically evaluate this expression 
to plot $m_u^*$ as a
function of $D$ in Fig.~\ref{Mu}. Inserting this 
$m_u^*$, function of $D$, into 
Eq.~(\ref{F_pi}), we also describe the renormalization scale $M_0$ as a 
function of $m_u$ and $D$. The $D$-dependence of $M_0$ is shown in 
Fig.~\ref{scale}. 
From Figs.~\ref{Mu} and \ref{scale}, we confirm that $m_u^*$ and $M_0$
in the 3-flavor NJL model have a behavior similar to 
that in the 2-flavor model \cite{Inagaki:2007dq}.

Next we  
would like to fix the remaining 3 parameters $\{G,\,\,K,\,\,m_s\}$. 
From Eqs.~(\ref{gap_u}), (\ref{gap_s}) and (\ref{m_K}) the coupling 
constants $G$ and $K$ are obtained as functions of $m_u, m_u^*$ and 
$m_s^*$,
\begin{align}
G(m_s^*)=&\frac{m_u^*-m_u}{4i{\rm tr}S^u} - \frac12
 K(m_s^*) i{\rm tr}S^s ,
\label{g4} \\
K(m_s^*)=&\frac{g(m_s^*)}{h(m_s^*)}, 
\label{g6}
\end{align}
where $g(m_s^*)$ and $h(m_s^*)$ are given by
\begin{align*}
g(m_s^*)=&-1+\frac{m_u^*-m_u}{2m_s^*}
\nonumber \\
 &+\frac{m_u^*-m_u}{2i{\rm tr}S^u}
  \left[ \frac{i{\rm tr}S^u}{m_u^*} 
  + \frac{i{\rm tr}S^s-i{\rm tr}S^u}{m_s^*} \right.
\nonumber \\
 &\left.+ \left\{ m_K^2 - (m_u^*-m_s^*)^2 \right\}
  I_{us}(m_K^2) \right] , \\
h(m_s^*)=&(i{\rm tr}S^s-i{\rm tr}S^u)\left[
  \frac{i{\rm tr}S^u}{m_u^*}+\frac{i{\rm tr}S^s}{m_s^*} \right.
\nonumber \\
 &\left. + \left\{ m_K^2 - (m_u^*-m_s^*)^2 \right\}
  I_{us}(m_K^2) \right].
\end{align*}
Substituting Eqs.~(\ref{g4}) and (\ref{g6}) 
into Eq.~(\ref{m_K}), 
we describe the current strange quark mass, $m_s$, as a function 
of $m_u, m_u^*$ and $m_s^*$.
Then the $\eta'$ meson mass (\ref{m_eta'}) is also expressed by a
function of $m_u, m_u^*$ and $m_s^*$.
From the expression for $m_{\eta'}$ and Eq.~(\ref{sol_gap_u})
we evaluate $m_s^*$ numerically.

\begin{figure}[!h]
  \begin{center}
    \includegraphics[height=2.0in,keepaspectratio]
    {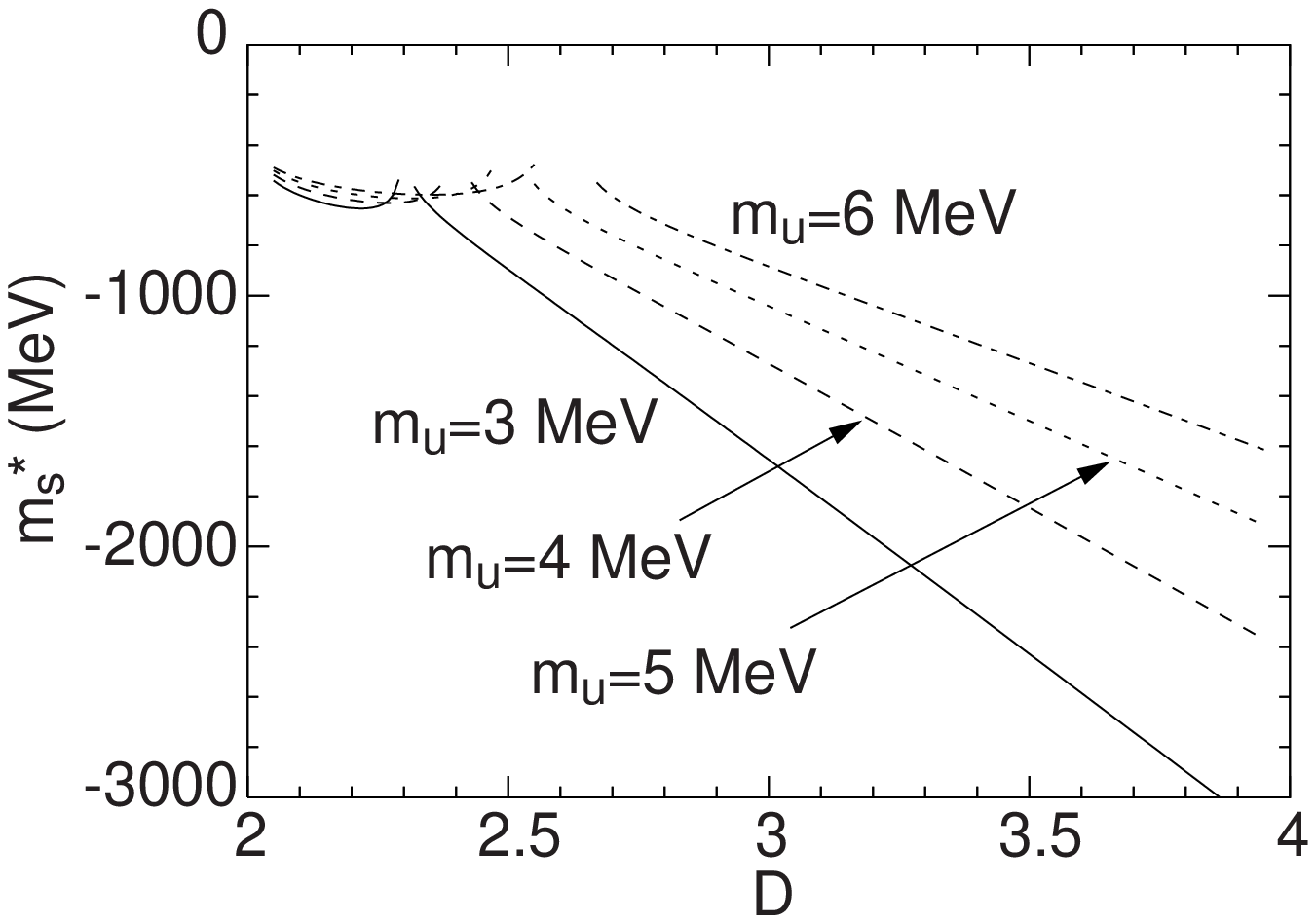} 
  \end{center}
  \vspace{-0.5cm}
  \caption{The solution of the gap equation $m_u^*$
   as a function of $D$ for $m_u=3,4,5$ and 6MeV.}
  \label{Ms}
\end{figure}

The $D$-dependence of $m_s^*$ is shown in Fig.~\ref{Ms}. We should note that 
qualitatively the solution for $m_s^*$ has a similar 
to $m_u^*$ behavior. Decreasing the dimension from four, we observe that the 
absolute value of $m_s^*$ goes down almost linearly.
A different tendency takes place near the dimension two, 
a discontinuity is observed at $D \simeq 2.5$. 
A physical solution for $m_s^*$ does not appear,
neither it is not seen for $m_u^*$ in Fig.~\ref{Mu}. 

\begin{figure}[!h]
  \begin{center}
    \includegraphics[height=2.0in,keepaspectratio]
    {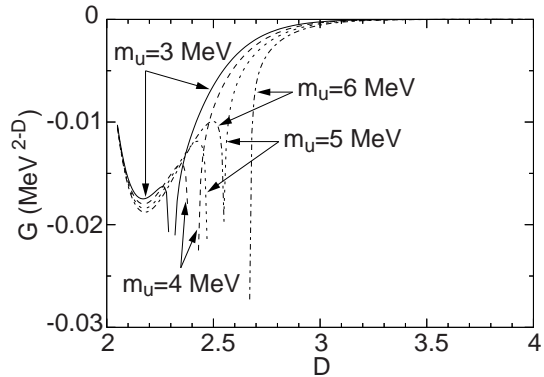} 
    \end{center}
  \vspace{-0.5cm}
  \caption{The 4-fermion coupling $G$ 
   as functions of $D$ for $m_u=3,4,5$ and $6$MeV.}
    \label{G}
\end{figure}

\begin{figure}[!h]
  \begin{center}
    \includegraphics[height=2.0in,keepaspectratio]
    {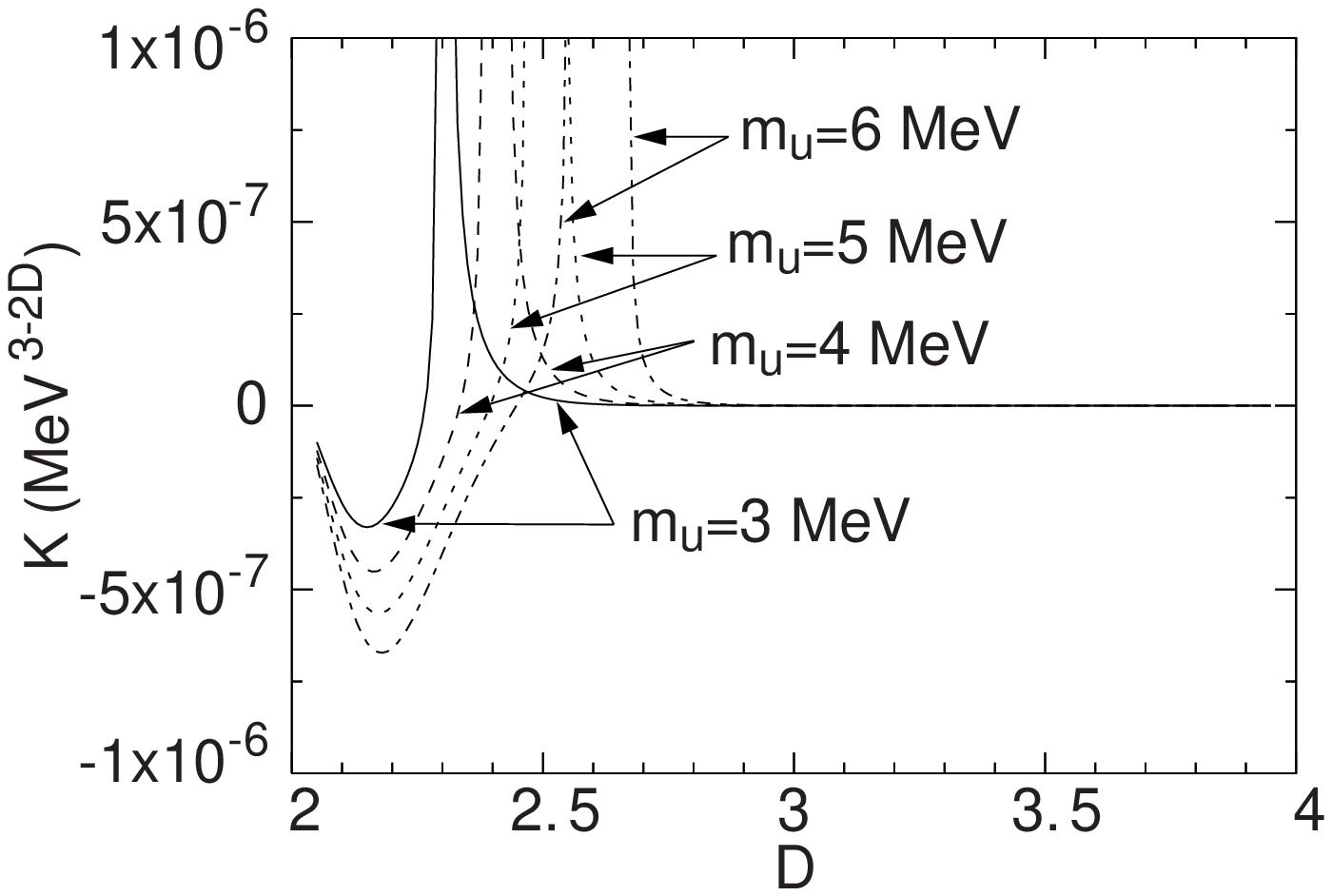} 
  \end{center}
  \vspace{-0.5cm}
  \caption{The 6-fermion coupling $K$ 
   as functions of $D$ for $m_u=3,4,5$ and $6$MeV.}
  \label{K}
\end{figure}

\begin{figure}[!h]
  \begin{center}
    \includegraphics[height=2.0in,keepaspectratio]
    {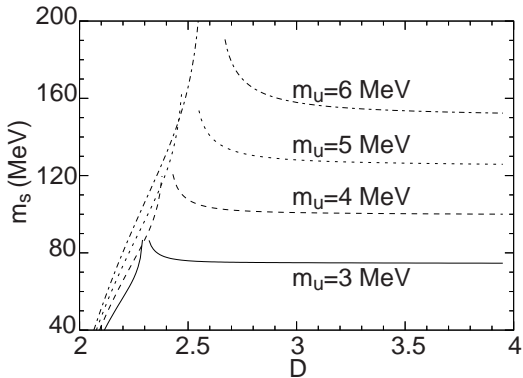} 
  \end{center}
  \vspace{-0.5cm}
  \caption{The strange quark mass $m_s$
   as a function of $D$ for $m_u=3,4,5$ and 6MeV.}
  \label{ms}
\end{figure}

Once $m_s^*$ is obtained, $G(m_s^*), K(m_s^*), m_s(m_s^*)$ are calculated by
inserting the solution for $m_s^*$. The numerical results for $G, K$ and $m_s$
are shown in Figs.~\ref{G}, \ref{K} and \ref{ms}, respectively. In Fig.~\ref{G} 
we observe a similar behavior of the four-fermion coupling, $G$, to  one in 
the 2-flavor NJL model \cite{Inagaki:2007dq}, while we do not have a consistent 
solution for the dimension where no solution is found for $m_s^*$. 
There is no analog of Fig.~\ref{K} in the 2-flavor case. 
In Fig.~\ref{K} we see that $K$ gets drastically large 
near the dimension where 
a consistent solution is lost. Finally in Fig.~\ref{ms}
 it is found that the value of the current strange quark mass $m_s$ is roughly 
constant for the dimension larger than that corresponding to the
discontinuity point, but it
falls down as $D$ decreases below the discontinuity point.

In Figs.\ \ref{Ms}, \ref{G}, \ref{K} and \ref{ms}, we observe a discontinuity
around $D\simeq2.5$ where the behavior drastically changed. It comes from the 
fact that the values of the couplings $G$ and $K$ become divergent when 
$m_s^*$ approaches $m_u^*$.
The self-energy $\Pi_5^{ii}(k^2=m_{\eta'}^2)$
is also divergent at $m_{\eta'}^2 = 4m_i^{*2}$.
Therefore $m_s^*$ has no physical solution around 
$m_s^* \simeq m_u^*\simeq m_{\eta'} /2$, as is numerically confirmed
in Fig.\ \ref{Ms}. This is the reason why the behavior of $m_s^*$ 
is different from that of $m_u^*$ near the dimension two. The constituent
 strange quark mass, $m_s^*$, can not develop a value smaller than 
$m_{\eta'} /2$. At the discontinuity  
the $\eta'$ meson propagator contains a imaginary part which corresponds to 
the decay width. 

\subsection{Physical quantities}
\begin{figure}[!h]
  \begin{center}
    \includegraphics[height=2.0in,keepaspectratio]
    {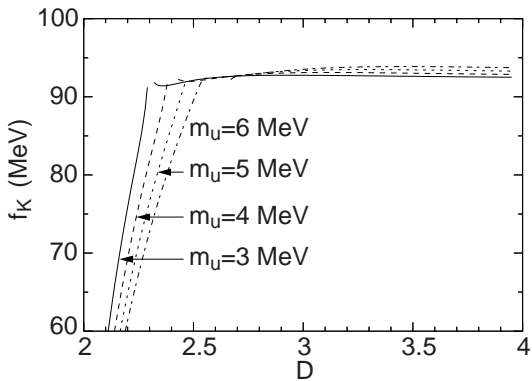} 
  \end{center}
  \vspace{-0.5cm}
  \caption{The pion decay constant $f_{K}$
   as a function of $D$ for $m_u=3,4,5$ and $6$MeV.}
  \label{fk}
\end{figure}

\begin{figure}[!h]
  \begin{center}
    \includegraphics[height=2.0in,keepaspectratio]
    {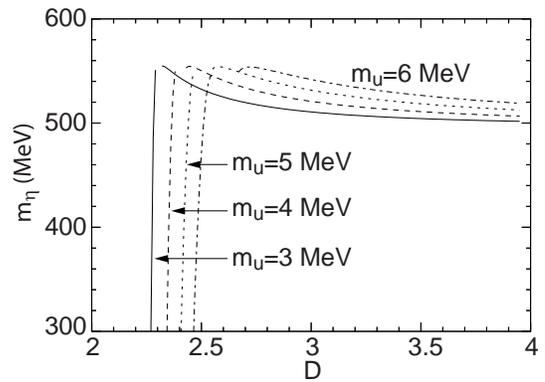} 
  \end{center}
  \vspace{-0.5cm}
  \caption{The $\eta$ mass $m_\eta$
   as a function of $D$ for $m_u=3,4,5$ and $6$MeV.}
  \label{m_eta_fig}
\end{figure}

\begin{figure}[!h]
  \begin{center}
    \includegraphics[height=2.0in,keepaspectratio]
    {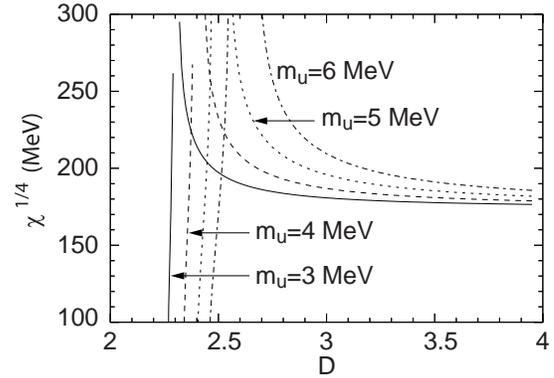} 
  \end{center}
  \vspace{-0.5cm}
  \caption{The topological susceptibility $\chi$
   as a function of $D$ for $m_u=3,4,5$ and 6MeV.}
  \label{topo}
\end{figure}

We have fixed the model parameters in the previous subsection except for the 
up quark mass $m_u$ and the dimension, $D$.
We are now ready for evaluating various meson properties as a function of $D$
for a fixed $m_u$.
Employing the obtained parameters and the chiral condensates, $m_u^*$ and
$m_s^*$, one can numerically calculate the kaon decay constant, $f_{K}$, the
$\eta$ meson mass, $m_\eta$, and the topological susceptibility, $\chi$.
The behavior of $f_{K},\ m_\eta$ and $\chi$ are displayed in Figs.~\ref{fk},
\ref{m_eta_fig} and \ref{topo}, respectively. These values are roughly constant
near four dimensions. They rapidly fall down as $D$ decreases near two
dimensions. We again observe a discontinuity around 
$m_s^* \simeq m_u^*\simeq m_{\eta'} /2$.
The topological susceptibility blows up around the discontinuity.


As is seen in Fig.\ \ref{fk}, the decay constant $f_{K}$ is smaller 
than the observed one, 110MeV, in the region, $2<D<4$. The model 
has to be improved to describe the Kaon decay. We can fit the $\eta$ meson
mass to $m_\eta=548$MeV by tuning the dimension of the fermion loop 
integrals, $D$. The topological susceptibility is calculated to
be $\chi^{1/4}=170\pm7,\ 174\pm7$MeV in lattice simulations 
\cite{Alles:1996mn} and $\chi^{1/4}=179$MeV in the Witten-Veneziano 
mass formula \cite{Witt:79, Vene:79}. We can also fit the 
topological susceptibility in the interval, $2<D<4$.
Consequently the dimension for the fermion loop integrals, $D$,
can be fixed to reproduce the $\eta$ meson mass or the topological 
susceptibility. We will discuss this matter in the next section
where we need to fit an additional parameter $D$.

\section{QUARK MASS DEPENDENCE}
In the cutoff regularization all the model parameters can be fixed
by fitting the physical quantities 
$\{m_{\pi},\, f_{\pi},\, m_{K},\, m_{\eta^{\prime}}\}$ for a given
$m_u$. 
In the dimensional regularization an an additional observable, $X'$, 
needs to be fitted due to one more parameter involved in this case. 
Therefore, there are five physical quantities, i.e. the set of them, 
$\{m_{\pi},\, f_{\pi},\, m_{K},\, m_{\eta^{\prime}},\, X'\}$, which are needed
to fit all the model parameters. In this section we 
consider two different observables for $X'$; one is the $\eta$ meson mass, 
$X'=m_{\eta}$,
another is the topological susceptibility, 
$X'=\chi$. By using the experimental/empirical values of $m_\eta$ 
and 
$\chi$ we can define two different parameter sets and evaluate the meson 
properties. Then some physical quantities are evaluated for 
$m_u=3$, $4$, $5$, $5.5$ and $6$MeV. We calculate
the physical quantities in the both regularizations to compare them with each other
for each given $m_u$.

\subsection{Physical quantities in the dimensional regularization}
To show the validity of the model as a low energy effective
theory of QCD here we evaluate parameters of the model 
to reproduce the input sets 
$\{m_{\pi},\, f_{\pi},\, m_{K},\, m_{\eta^{\prime}}, m_\eta\}$ or
$\{m_{\pi},\, f_{\pi},\, m_{K},\, m_{\eta^{\prime}}, \chi\}$ in the
dimensional regularization and discuss the other (output) physical
quantities.
There exist two solutions to reproduce $m_\eta=548$MeV for 
$m_u\lesssim 5$MeV. In Table \ref{phys_dim1} we show
$m_s,\ f_K,\ \chi,\ \langle \bar{u}u \rangle_r$ under the condition
that $m_\eta=548$MeV.
The subscript $r$ in $\langle \bar{u}u \rangle_r$
stands for that the quantity is renormalized,
$\langle \bar{u}u \rangle_r = \langle \bar{u}u \rangle M_0^{4-D}$.
The last lines in Table \ref{phys_dim1} and \ref{phys_dim2} show the experimental/empirical
values. The light quark masses are evaluated at 1GeV \cite{PDG}.
The empirical value of $\langle \bar{u}u \rangle \simeq
\langle \bar{s}s \rangle$ is evaluated by using the
Gell-Mann-Oakes-Renner relation \cite{GellMann:1968rz}.
We observe that
the topological susceptibility $\chi$ is always larger than 
the one in the lattice simulation and the Witten-Veneziano mass 
formula.

In the Tables \ref{phys_dim2} and \ref{phys_dim3} we fix the parameters of the model
so that to reproduce $\chi^{1/4}=170$ or $179$MeV and then we
calculate $m_s,\ f_K,\ m_\eta,\ \langle \bar{u}u \rangle_r$ for
$m_u=3, 4, 5, 5.5, 6$MeV.
As is shown in Fig.~\ref{topo}, we find two solutions in both 
sides of the discontinuity to satisfy $\chi^{1/4}=179$MeV for 
$m_u\lesssim 4$MeV. 
In any case, we obtain the value for the $\eta$ meson mass
which is smaller than the observed one, $m_\eta = 548$MeV.

\begin{table} [!h]
\caption{Physical quantities 
in the dimensional regularization
(in units of MeV, except for $D$), 
$m_\eta$ is fixed at 548MeV. The last line shows
the experimental/empirical values.}
\label{phys_dim1}
\begin{ruledtabular}
\begin{tabular}{cccccc}
$m_u$ & $m_s$ & $f_K$ & $\chi^{1/4}$ &  
$-\langle \bar{u}u \rangle_r^{1/3}$ & $D$ \\
\hline 
3.0 & 84.9 & 90.8 & 244  & 301 & 2.289 \\
3.0 & 79.0 & 91.4 & 225  & 301 & 2.372 \\
4.0 & 118  & 91.4 & 254  & 274 & 2.378 \\
4.0 & 106  & 92.1 & 225  & 273 & 2.522 \\
5.0 & 156  & 92.0 & 275  & 254 & 2.465 \\
5.0 & 134  & 92.7 & 224  & 254 & 2.687 \\
5.5 & 148  & 92.9 & 224  & 246 & 2.775 \\
6.0 & 162  & 93.2 & 224  & 239 & 2.867 \\
\hline
$3.4-6.8$ & $94.5-176$ & $110$ & $170-179$ & $228-287$ &
\end{tabular}
\end{ruledtabular}
\end{table}

\begin{table} [!h]
\caption{Physical quantities 
in the dimensional regularization
(in units of MeV, except for $D$), 
$\chi^{1/4}$ is fixed at 170MeV. The last line shows
the experimental/empirical values.}  
\label{phys_dim2}
\begin{ruledtabular}
\begin{tabular}{cccccc}
$m_u$ & $m_s$ & $f_K$ & $m_\eta$ &  
$-\langle \bar{u}u \rangle_r^{1/3}$ & $D$ \\
\hline 
3.0 & 77.1 & 88.4 & 481  & 302 & 2.280 \\
4.0 & 105  & 88.8 & 478  & 274 & 2.360 \\
5.0 & 134  & 89.2 & 475  & 255 & 2.434 \\
5.5 & 150  & 89.3 & 473  & 247 & 2.467 \\
6.0 & 166  & 89.5 & 471  & 240 & 2.500 \\
\hline
$3.4-6.8$ & $94.5-176$ & $110$ & $548$  & $228-287$ &
\end{tabular}
\end{ruledtabular}
\end{table}

\begin{table} [!h]
\caption{Physical quantities for the dimensional regularization
(in units of MeV, except for $D$), 
$\chi^{1/4}$ is fixed at 179MeV.}  
\label{phys_dim3}
\begin{ruledtabular}
\begin{tabular}{cccccc}
$m_u$ & $m_s$ & $f_K$ & $m_\eta$ &  
$-\langle \bar{u}u \rangle_r^{1/3}$ & $D$ \\
\hline 
3.0 & 78.0 & 88.7 & 498  & 301 & 2.281 \\
3.0 & 74.8 & 92.7 & 507  & 300 & 3.222 \\
4.0 & 106  & 89.1 & 495  & 274 & 2.363 \\
4.0 & 100  & 92.9 & 507  & 272 & 3.895 \\
5.0 & 136  & 89.5 & 491  & 255 & 2.438 \\
5.5 & 152  & 89.7 & 489  & 247 & 2.472 \\
6.0 & 168  & 89.9 & 487  & 240 & 2.505
\end{tabular}
\end{ruledtabular}
\end{table}

\subsection{Physical quantities in the cutoff regularization}
In this section we fix the model parameters 
to reproduce the measured values of Eq.(\ref{input}) in the cutoff regularization for
a given $m_u$. The quantities, $\{ m_s, G, K, \Lambda \}$, 
are found following the
same procedure as in the previous section. In the Table \ref{para_cut} we show 
the parameters, $G, K$ and $\Lambda$ for $m_u=3$, $4$, $5$, $5.5$ and 
$5.87$MeV.
It is observed that the cutoff scale $\Lambda$ decreases as $m_u$ 
increases. No solution is found to simultaneously satisfy
Eqs.\ (\ref{m_pi}) and (\ref{F_pi}) for $m_u \gtrsim 5.87$MeV.
In Table \ref{phys_cut} 
the quantities, 
$m_\eta,\ \chi,\ m_s,\ \langle \bar{u}u \rangle$, are shown. In the cutoff
regularization the kaon decay constant, $f_K$, is consistent with
the observed value for a smaller $m_u$, while the $\eta$ meson
mass and the topological susceptibility are smaller than the 
experimental/empirical values. We obtain only smaller 
$\eta$ meson mass than the experimental one, 
similarly to the situation in the dimensional regularization with 
fixed topological susceptibility $\chi$.

\begin{table} [!h]
\caption{Parameter list for cutoff scale $\Lambda$, 4-fermion coupling 
$G$ and 6-fermion coupling $K$.}
\label{para_cut}
\begin{ruledtabular}
\begin{tabular}{cccc}
$m_u$(MeV) & $\Lambda$(MeV) & $G\Lambda^2$ & 
$K\Lambda^5$ \\
\hline 
3.0 & 960 & 1.55 & 8.34 \\
4.0 & 797 & 1.60 & 8.38 \\
5.0 & 682 & 1.71 & 8.77 \\
5.5 & 630 & 1.81 & 9.17 \\
5.87 & 580 & 2.09 & 10.1 
\end{tabular}
\end{ruledtabular}
\end{table}

\begin{table} [!h]
\caption{Physical quantities in the cutoff regularization
(in units of MeV).}
\label{phys_cut}
\begin{ruledtabular}
\begin{tabular}{cccccc}
$m_u$ &  $m_s$ & $f_K$  & $m_\eta$ & $\chi^{1/4}$ & 
$-\langle \bar{u}u \rangle^{1/3} $ \\
\hline 
3.0 & 89.5 & 113  & 451 & 160 & 301 \\
4.0 & 110  & 107  & 457 & 158 & 273 \\
5.0 & 128  & 101  & 473 & 160 & 253 \\
5.5 & 136  & 97.3 & 482 & 163 & 245 \\
5.87 & 139 & 93.3  & 501 & 172 & 240 
\end{tabular}
\end{ruledtabular}
\end{table}

Thus none of the regularizations can
describe all of the nonet meson properties. We 
can fit the $\eta$ meson mass or the topological susceptibility in 
the dimensional regularization but we obtain only a smaller kaon decay 
constant in either case. 
In the Pauli-Villars regularization it has been 
found that 
$m_u=m_d=2.7$MeV, $m_s=92$MeV, $f_K=131$MeV and $m_\eta=526$MeV 
where the input parameters $m_\pi,\ m_K,\ f_\pi$ and $m_{\eta'}$ have been used 
\cite{Osipov:2004mn}.

\section{CONCLUSION}
We studied nonet meson properties in the three-flavor NJL 
model with the dimensional and sharp cutoff regularizations in the
leading order of the $1/N_c$ expansion.
We employed $m_u,\ m_\pi,\ m_K,\ f_\pi,\ m_{\eta'}$ as input parameters
and fix the model parameters, $m_s,\ G,\ K,\ M_0$, and 
$m_s,\ G,\ K,\ \Lambda$ in the dimensional and cutoff 
regularizations, respectively. In the case of the dimensional 
regularization the dimension, $D$, is still 
a free parameter. Thus we 
evaluate the kaon decay constant, the $\eta$ meson mass and the 
topological susceptibility as a function of $D$.

The constituent up-quark mass, $m_u^*$, and the renormalization scale
$M_0$ 
behave in a similar to the two-flavor case way. 
No consistent solution for $m_s^*$ is found and the effective coupling,
$G$ and $K$, are divergent around 
$m_s^*\simeq m_u^*\simeq m_{\eta'}/2$.
Below this region $\eta'$ meson propagator develops an imaginary part.
The kaon decay constant, $f_K$, is smaller than the observed value, 
110MeV, for $2<D<4$.

We found that $m_s,\ f_K,\ m_\eta,\ \chi$ are almost constant near
four dimensions. Fitting the parameters at lower dimension and then
taking the four dimensional limit, obtained values for 
$m_s,\ f_K,\ m_\eta,\ \chi$ are not divergent. The results 
do not depend on the regularization parameter. It bring us an 
interesting idea of making a regularization independent prediction 
in the NJL model.
  
We have evaluated the physical quantities, $m_s,\ f_K,\ \chi$,
$m_\eta$ and $\langle \bar{u}u \rangle$, in both the dimensional 
and the cutoff regularizations for a fixed $m_u$. In the 
dimensional regularization the topological susceptibility,
$\chi$, develops a larger value than the one obtained in the
lattice simulation and Witten-Veneziano mass formula for the 
fixed $m_\eta$. On the other hand, the $\eta$ meson acquires a 
smaller mass than the observed one for the fixed $\chi$. 

In the cutoff regularization the cutoff scale significantly 
depends on $m_u$; the cutoff $\Lambda$ increases with decreasing
$m_u$. It is interesting to note that the coupling constants
become smaller when one takes the larger cutoff, which is
consistent with the renormalization group argument where the
coupling strength becomes smaller with increasing the energy-scale.
This cutoff effect is also confirmed numerically in the NJL model
through changing the cutoff in the temporal direction \cite{Chen:2009mv}.

In the Tables. \ref{phys_dim2}, \ref{phys_dim3} and \ref{phys_cut}
some difficulty is seen in tuning 
$f_K,\ m_\eta$ and $\chi$ simultaneously.
Thus, this paper shows that the model based on the Lagrangian (\ref{LNJL})
is not satisfactory in either of regularizations. It teaches us that new terms
should be added to the Lagrangian to improve the model at least in one of
the regularizations. Especially,
we are interested in including vector type
and multi-fermion interactions 
and hope to report on the problem in future.

\begin{acknowledgments}
The authors would like to thank Y.~Hoshino and Y.~Kitadono for fruitful discussions.
HK is supported by the grant NSC-99-2811-M-033-017 from
National Science Council (NSC) of Taiwan.
Discussions during 2009 International ``Workshop on Strong Coupling 
Gauge Theories in LHC Era''(SCGT 09) and
the YIPQS international workshop on ``New Frontiers in QCD 2010'', 
were useful to complete this work.
\end{acknowledgments}

\appendix
\section{Integrals}
In this section, we demonstrate several integrals both in the cutoff and
dimensional regularization schemes which are required 
in the section \ref{njl_model} and \ref{meson}.

\subsection{The cutoff regularization}
The chiral condensates $i\,{\rm tr}S^i$ in Eq.(\ref{trace}) are important
quantities because they concern gap equations. 
They take the following form after the integration is performed,
\begin{eqnarray}
 i\, \trsc S^i 
 &=& \frac{N_c}{2\pi^2} m_i^* \left\{
  \Lambda \sqrt{\Lambda^2 + m_i^{*2}} \right. \nonumber \\ 
 &&\left. - m_i^{*2} 
  \ln \frac{\Lambda + \sqrt{\Lambda^2 + m_i^{*2}}}
  {\sqrt{m_i^{*2}}} \right\}.
\label{trS_cut}
\end{eqnarray}
The next integral to carry out is $I_{ij}(k^2)$ in Eq.(\ref{I_ij}) which
determines self-energies of mesons, $\Pi_P$, Eq.(\ref{Pi_P}). It
takes the form
\begin{eqnarray}
I_{ij}
 &=& \frac{N_c}{8\pi^2}\int_0^1 dx
  \left\{ -\frac{\Lambda}{\sqrt{\Lambda^2+L_{ij}(k^2)}} \right.
  \nonumber \\
 &&\left. +\ln  \frac{\Lambda+\sqrt{\Lambda^2+L_{ij}(k^2)}}
   {|{\sqrt{L_{ij}(k^2)}}|} \right\},
  \label{I_ij_cut}
\end{eqnarray}
where
\begin{equation}
L_{ij}(k^2)=m_i^{*2} - (m_i^{*2}-m_j^{*2})x - k^2x(1-x) .
\label{L_ij}
\end{equation}
Finally, Eqs.\ (\ref{F_K}) and (\ref{J_us}) involved in the derivation of the kaon decay
constant are calculated as
\begin{align}
f_K^2=
 &\frac{4}{J_{us}(0)} \left[m_u^* I_{us}(0)
  - \frac{N_c}{8\pi^2}(m_s^*-m_u^*) \right. 
  \nonumber \\
 & \times \int_0^1 dx x \left\{\frac{\Lambda}
  {\sqrt{\Lambda^2+L_{us}(0)}} \right. \\
 &\left. \left. - \ln\frac{\Lambda+\sqrt{\Lambda^2+L_{us}(0)}}
  {\sqrt{L_{us}(0)}} \right\} \right]^2 , \nonumber \\
J_{us}(k^2)=
 & I_{us}(k^2)-\frac{N_c}{16\pi^2}
   \Lambda^3 (m_s^*-m_u^*)^2 \nonumber \\
 &\times\int_0^1 dx \frac{x(1-x)}
  {L_{us}(k^2)[\Lambda^2+L_{us}(k^2)]^{3/2}} .
\end{align}

\subsection{The dimensional regularization}
The corresponding integrals in the dimensional regularization 
are performed as
\begin{align}
 i\, \trsc S^i 
 &= \frac{N_c}{(2\pi)^{D/2}} 
  \Gamma \left(1- \frac{D}{2} \right) 
   m_i^* (m_i^{*2})^{D/2-1} ,
  \label{trS_dim} \\
f_K^2 
 &= \frac{2^{D/2}}{M_0^{D-4} J_{us}(0)} \left[m_u^* I_{us}(0)
   +\frac{N_c}{(4\pi)^{D/2}}\Gamma\left( 2-\frac{D}{2}
    \right) \right. \nonumber\\
 &\left. \quad \times
  (m_s^*-m_u^*)\int_0^1 dx xL_{us}(0)^{D/2-2} \right]^2 , \\
J_{us}
 &= I_{us}(k^2) - \frac{N_c}{(4\pi)^{D/2}}
    \Gamma\left( 3-\frac{D}{2} \right) (m_s^*-m_u^*)^2 
  \nonumber \\
 &\quad \times \int_0^1 dx x(1-x)
  L_{us}(k^2)^{D/2-3}. 
\end{align}
Regarding $I_{ij}$, we have different expressions depending on the value
of $k^2$;
\begin{align}
I_{ij}
 =
  \frac{N_c}{(4\pi)^{D/2}} \Gamma \left(2-\frac{D}{2} \right)
   \int_0^1 dx L_{ij}^{D/2-2}(k^2) ,
 \label{I_ij_dim1}
\end{align}
for $(m_i^*-m_j^*)^2 < k^2 < (m_i^*+m_j^*)^2$ and $k^2=0$,
\begin{align}
&I_{ij}(k^2) =\frac{N_c}{(4\pi)^{D/2}} 
 \Gamma \left(2-\frac{D}{2} \right) 
 \left[ \frac2{D-2} \nu_{ij}^{D/2-2} \right. 
\nonumber \\
& \times \left\{ a_{-}^{D/2-1} 
 F\left(2-\frac{D}{2},\frac{D}{2}-1,\frac{D}{2};
 -\frac{a_{-}}{a_{+}-a_{-}} \right) 
\right.
\nonumber \\
& + \left. (1-a_{+})^{D/2-1} 
 F\left(2-\frac{D}{2},\frac{D}{2}-1,\frac{D}{2};
 -\frac{1-a_{+}}{a_{+}-a_{-}} \right)
 \right\}
\nonumber \\
& + \left. e^{i\pi(2-D/2)}(k^2)^{1-D/2}\nu_{ij}^{D-3}
 B\left(\frac{D}{2}-1,\frac{D}{2}-1\right) \right],
\label{I_ij_dim2}
\end{align}
for $0<k^2<(m_i^*-m_j^*)^2,\ (m_i^*+m_j^*)^2<k^2$, respectively.
$F$ denotes the hypergeometric function.
In Eq.\ (\ref{I_ij_dim2}) we introduce the quantities $a_{\pm}$
and $\nu_{ij}$ which are defined by
\begin{align*}
a_{\pm}&=\frac{k^2+(m_i^{*2}-m_j^{*2})
 \pm\nu_{ij}}{2k^2} , \\
\nu_{ij}&=\sqrt{k^4-2k^2(m_i^{*2}+m_j^{*2})+
 (m_i^{*2}-m_j^{*2})^2} .
\end{align*}


\end{document}